\newlength{\fl} 

\documentstyle[aps,aipbook,floats,epsf]{revtex}
\begin{document}
\renewcommand{\floatpagefraction}{.8}
\title{Status of Lattice QCD\thanks{Talk presented at the Lepton-Photon 
   Symposium, Cornell University, Aug. 10-15, 1993}}

\author{Paul B. Mackenzie}
\address{Fermi National Accelerator Laboratory
\\
Batavia, Illinois 60510
}

\maketitle

\begin{abstract}
Significant progress has recently been achieved
in the lattice gauge theory
 calculations required for extracting the fundamental parameters
of the standard model from experiment.
Recent lattice determinations of such  quantities as
 the kaon $B$ parameter,
the mass of the $b$ quark, and the strong coupling constant
have produced results and uncertainties
as good or better than the best conventional determinations.
Many other calculations crucial to extracting the fundamental
parameters of the standard model from experimental data
 are undergoing very active development.
I review the status of such  applications of lattice QCD
to standard model phenomenology,
and discuss the prospects for the near future.
\end{abstract}

\section*{Introduction}
Our only existing experimental clues about the theory that lies 
beyond the standard model 
are the apparently arbitrary fundamental parameters of the
standard model.
 The only experiments guaranteed to determine the origin of
electroweak symmetry breaking and thus new clues into
beyond the standard model physics were to have been performed at the
SSC.  
The fundamental parameters of the standard model may prove
to be our only window onto beyond the standard model physics
for some time,
unless we get lucky with a  lower energy accelerator.
The nonperturbative calculations  which allow
the extraction of these parameters from experiment will therefore
take on increasing importance over the next few years.

The last few years have seen significant progress in 
some of these calculations with
lattice QCD.
Some of the simplest ones have now been completed with
first attempts at quantitative estimation of all uncertainties.
 Lattice calculations of the mass of the $b$ quark,
$m_b$, and the kaon $B$ parameter, $B_K$,
are now believed by their authors to be more accurate than the best
conventional phenomenological determinations of these quantities.
 Determinations of the strong coupling constant, $\alpha_s$,
are now of comparable quality to the best conventional determinations
and will soon be significantly better.
Many other calculations crucial to standard model phenomenology are undergoing
 rapid development.

The foundations of the current advances were laid around 1980
with the
development of explicit expressions for hadron masses and other hadronic
quantities by Weingarten and by Parisi and their collaborators.~\cite{Parisi}
The recent developments have arisen in part because of dramatic developments
in machine and algorithmic technology that will not be reviewed 
here.~\cite{confs}
The computers on which these calculations have been performed
are around 10,000 times as powerful as the Vaxes on which the first hadron
spectrum calculations were performed around 1980. 
Likewise, the speeds of 
 algorithms for the inclusion of sea quark loops from first 
principles (algorithms for ``unquenched'' calculations) have 
increased by an even larger factor, one which is hard to measure
because of the extreme slowness of the original algorithms.
Numerous other methodological and technical
improvements have also contributed to the
reliability of the calculations.

These developments  have occurred
 also because of an improved perspective regarding which
lattice calculations are easiest to perform reliably, and are most useful to 
particle physics.
Although the desire to understand  the physics of nuclear 
energy levels was the initial spur to the study of strong interactions,
the calculation of the energy levels of the uranium nucleus
is not currently seen as the most promising test of QCD, or of standard
model or beyond the standard model physics.
Likewise, although the ability to calculate the proton mass from first
principles seemed like a Holy Grail when lattice gauge theory was 
invented 20 years ago, the mass of the proton and the rest of the
light hadron spectrum is not the only
or even the most important application of lattice gauge theory.
The  pseudoscalar mesons $\pi$, $K$, $D$, and $B$, and the
quarkonia (the $\psi$'s and $\Upsilon$'s) are significantly simpler
than the proton and other hadrons, as
 will be discussed.  They will provide good tests of lattice methods
and significant information about the standard model before calculations
of the proton mass do.

\section*{QCD}

\subsection*{QCD Phenomenology}

\subsubsection*{The $\psi$ and $\Upsilon$ Systems}
The simplest hadrons to investigate on the lattice are the  
$\psi$ and $\Upsilon$ systems.  
Quarkonia are smaller than most hadrons, resulting in smaller finite
volume errors.
 Propagators for heavy quarks can be calculated much more
rapidly than light quark propagators, and no extrapolation down to the
physical quark mass is required as it is for light quarks.
 These facts have been particularly emphasized in Ref.~\cite{ThackerLepage}.

Quarkonia 
have received relatively little attention from lattice theorists until 
recently.  The reason may be that they have been well understood
for a long time on the basis of nonrelativistic potential models
\cite{eetal}, which become rigorous predictions of QCD in a well-defined
limit, $m_Q\rightarrow\infty$.
This fact should, to the contrary, place them among the most interesting 
and important of current lattice
calculations because of the possibility of using nonrelativistic methods
and reasoning to:
\begin{itemize}
 \item guide physics expectations, and 
 \item monitor the accuracy of approximations (finite lattice spacing $a$,
   finite volume $V$, quenched) and  make corrections.
\end{itemize}

For example, the nonrelativistic picture tells us that the hyperfine 
splitting in quarkonia is a short distance quantity, sensitive to  
$\overline{\psi}\sigma \cdot B \psi$,  the dimension five operator which
is the leading operator which must be added to the action to correct for
finite lattice spacing errors.  
This quantity is useful in testing and fine tuning the approximations
used in lattice calculations.
A spin-averaged quantity like the
1P--1S splitting should be insensitive to these leading
finite lattice spacing errors.  It is also insensitive to the precise value
of the quark mass used, since it is almost the same for the $\psi$ and
$\Upsilon$ systems.  
It is  a likely candidate to yield information about particle physics,
as in the extraction of the strong coupling constant, to be discussed below.

\begin{figure}
\epsfxsize=   3in
\centerline{\hbox{\epsfbox{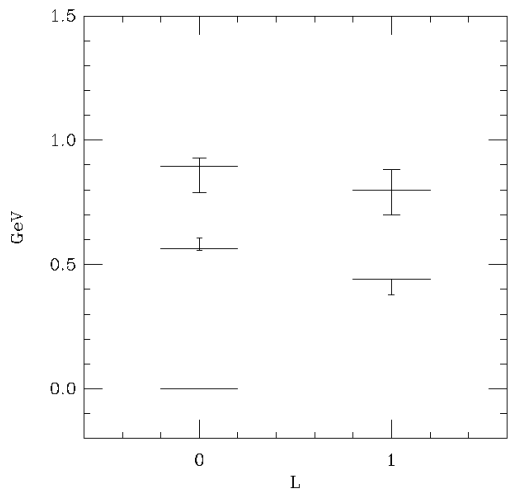}}}
\caption{The mass spectrum of the $\Upsilon$ system,
   $L=0$ states (1S, 2S, and 3S) and $L=1$ states   (1P and 2P).
   Mass difference in GeV with the ground state is shown.
  Solid lines are experiment.
}
 \label{upspec}
\end{figure}

The most extensive investigation of finite lattice spacing errors
which has yet been performed in a phenomenological calculation has recently
been completed by the NRQCD collaboration for the
$\psi$ and $\Upsilon$ systems.  
They use the formalism of Nonrelativistic QCD \cite{ThackerLepage},
a discretized version of the nonrelativistic expansion of the quark action.
In previous work reported in Ref.\cite{lep91}, 
coefficients of  operators for finite lattice spacing and
nonrelativistic corrections were evaluated at tree level.
Corrections were then determined by evaluating the operator
expectation values in potential model wave functions.
This year, these expected corrections were verified from first principles
by including the required operators directly in the lattice calculations,
with coefficients evaluated to one loop.  \cite{NRQ93}
The resulting spectrum for the $\Upsilon$ is shown in Fig.~\ref{upspec}.
The mass of the $\Upsilon$ was used as an input (to fix the
quark mass), and the overall
energy scale (the lattice spacing in physical units) was chosen to
obtaining the best fit to the remaining masses.

\begin{figure}
\epsfxsize=   \textwidth 
\epsfbox{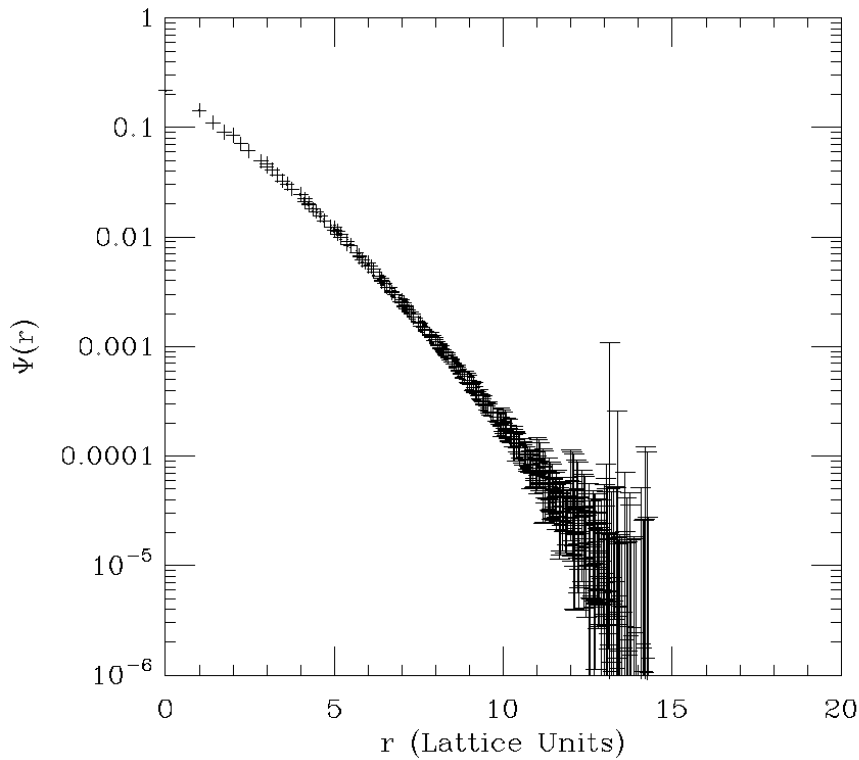}
\caption{The wave function $\Psi(r)$ of the $\Upsilon$ meson 
 as a function of $r$ in lattice units.}
\label{upwf}
\end{figure}

Fig.~\ref{upwf} shows the Coulomb gauge wave function, calculated 
on a $24^4$ lattice with Wilson fermions. \cite{elk93}  
Unlike a QED-like pure Coulomb
potential which would produce a wave function with the
form $\Psi(r)\propto\exp(-\alpha m r)$, QCD produces 
nonrelativistic bound states with wave 
functions that fall more slowly at short distances (smaller effective
$\alpha$ at short distances) and faster at long distances (as
expected from confinement).
This wave function was calculated in the quenched approximation, which
has slightly too much asymptotic freedom due to the absence of light
quark loops.  It would not be surprising to find that the same
 calculations repeated in the full theory showed slightly less 
concavity in the wave function,
perhaps of order 20\% less ($\approx \beta_0^{nf=3}/\beta_0^{nf=0}=9/11$).
The effects of finite boundary conditions can be seen half way
across the lattice, at $r=12$
in lattice units.

\subsubsection*{The Light Hadron Spectrum}

Calculations involving light quarks are significantly more difficult
than those with only heavy quarks.  
One cannot estimate as accurately in advance what order of correction
operators must be added to the action to achieve a certain accuracy in
finite lattice spacing errors,
or what volume must be used to reduce finite volume errors to a
negligible level.
These things must be determined to a much greater degree by painstaking
experimentation.
Light quark propagators are much more costly computationally.
Effects of light quark loops are likely to be  more complicated than
for quarkonia.

Current algorithms for calculating light quark algorithms fail when 
the light quark mass $m_l\equiv (m_u+m_d)/2$
is reduced toward its physical value.
The extrapolation is likely to be reasonably straightforward
for such quantities as $m_\pi$ and $f_\pi$ whose chiral behavior
is well understood.
The masses of particles like the $\rho$ which become unstable in the
small $m$, large $V$ limit clearly require special care in 
extrapolating to $m_l\rightarrow 0$. \cite{lue-vol}
(The correlation functions from which $M_\rho$ is determined become
dominated by the two pion cut, rather than the $\rho$ pole in
the physical limit.)
The proton mass is also known to have much larger nonlinear corrections
in chiral perturbation theory 
than  $m_\pi$ and $f_\pi$ do when $m_l$ is raised above $m_s$.~\cite{xipro}
Analogous effects in $M_P$ occur in the quenched approximation with different 
coefficients.  In addition, there are  indications in quenched chiral
perturbation theory calculations of pathologies as
$m_l\rightarrow 0$, which so far have not been reconciled with
numerical results. \cite{qxipt}

\begin{figure}
\epsfxsize=   \textwidth 
\epsfbox{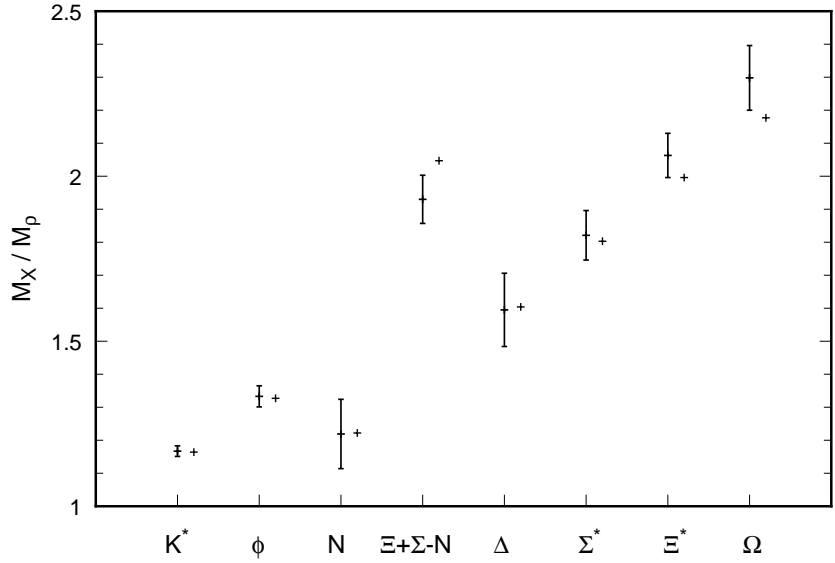}
\caption{The spectrum of the light hadrons in the quenched approximation,
  extrapolated to zero lattice spacing, infinite volume, and 
  physical quark mass. 
  Error bars do not include
  uncertainties due to the quenched approximation 
  or to finite lattice spacing.
  + denotes experiment.
}\label{fig:spectrum}
\end{figure}

The most systematic attempt so far at calculating the hadron spectrum
in the quenched approximation appeared this year from the GF11
collaboration.~\cite{gf11}
The calculation was performed at several values each of lattice 
spacing, volume, and quark mass, with the results extrapolated to
the physical values of each.
The resulting spectrum is shown in Fig.~\ref{fig:spectrum}.
$M_\pi$ and $M_K$ have been used as inputs to set the quark masses.
The error bars include statistics, finite volume, and $m_l\rightarrow 0$
extrapolations (assuming linear behavior in $m_l$).
They do not include uncertainties due the finite lattice spacing
extrapolation
and to the quenched approximation.
The dominant finite lattice spacing errors can be removed by adding a single
correction operator to the action.  
A calculation with a tree level improved $O(a)$ corrected action has been
performed at a single small lattice spacing, with results that appear
to be consistent with these extrapolated results (though with larger
statistical errors).~\cite{ukq93}

\begin{figure} 
\vspace*{\fl}
\caption{The pion decay constant  calculated in full QCD with two light 
  flavors, as a function of the quark mass.  Light hadron results must at
  present be extrapolated to the physical quark mass because algorithms
  fail when the mass becomes too small.
}\label{fig:fpi}
\end{figure}

Adding light quark loops to the calculation is much more expensive.
The state of uncertainty analysis is therefore somewhat less advanced
than for the quenched theory.
Even more than for quenched calculations, algorithms begin to fail
as the quark mass is reduced.  The light quark mass can be
reduced only to around $0.4\ m_s$;  results must then be extrapolated to the
physical limit.
Fig.~\ref{fig:fpi} shows an unquenched calculation of $f_\pi$ \cite{Fuk93},
which may be the simplest light hadron quantity to obtain other than
the pion mass.
 Chiral perturbation theory leads us to expect it to have a 
small and smooth extrapolation to the chiral limit,
compared with $M_\rho$ and $M_P$.
The calculation used two light flavors of quarks in the sea,
the lattice quark masses of 0.02 and 0.01
correspond roughly to $m_s$ and $m_s/2$. 
The size of the extrapolation from $m_l\approx m_s/2$ 
($M_\pi^2\approx M_K^2$) is consistent with expectations based
on the experimental value of $f_K^2/f_\pi^2$.
Some sources of error have been carefully checked:
consistency between values of $f_\pi$ obtained from various operators has been
tested,
the effects of finite volume on $M_\pi$ have been tested to be under
1\% (finite volume effects on $M_\rho$ and $M_P$ are much larger).
Yet to be checked are the effects of finite volume on $f_\pi$ itself,
the agreement of  $f_\pi$ as a function of $m_l$ with chiral perturbation
theory, and the effects of finite lattice spacing.

The algorithmic restriction to unphysically large light quark masses seems
likely to be with us for a while.
As the various sources of uncertainty in unquenched calculations gradually
become better understood, 
one of the important questions for consumers of lattice calculations will
become:
which quantities have the smoothest and best understood extrapolations
from $m_l \approx m_s/2$ down to the physical light mass  limit?

\subsection*{The Fundamental Parameters of QCD}
From the standpoint of standard model physics, one of the most 
crucial results of hadron spectrum calculations is the determination
of the fundamental parameters of QCD:
the strong coupling constant, $\alpha_s$, and the quark masses.
Such calculations have two elements.
 First, one calculates a measurable
dimensionful quantity such as $f_\pi$ or a level splitting in the $\psi$
or $\Upsilon$ system, to set the lattice energy scale in physical units.
 This appears to be the least important source of uncertainty 
in such calculations.
Second, one determines the physical coupling at short distances.
This may be done
either by a) relating the bare lattice parameter to a conventional definition
($\overline{MS}$, for example) via renormalization group improved,
mean field improved perturbation theory, or better still, by b) ignoring
the bare lattice parameters (which are often somewhat pathological
compared with physically defined parameters) and extracting the physical
parameters nonperturbatively from short distance lattice calculations.

\begin{figure} 
\hbox{\epsfxsize= 2.2 in\epsfbox{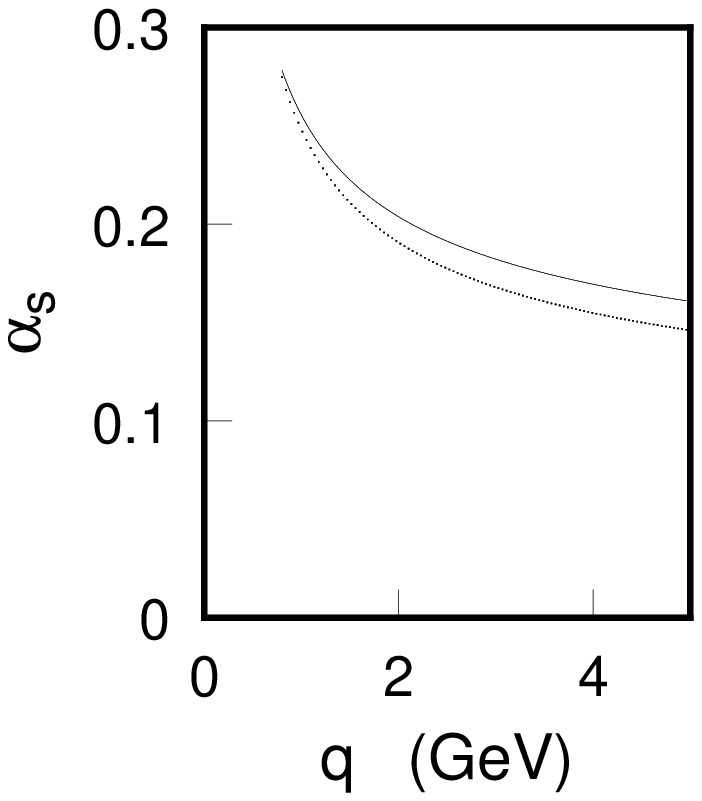}\epsfxsize= 2.2in \epsfbox{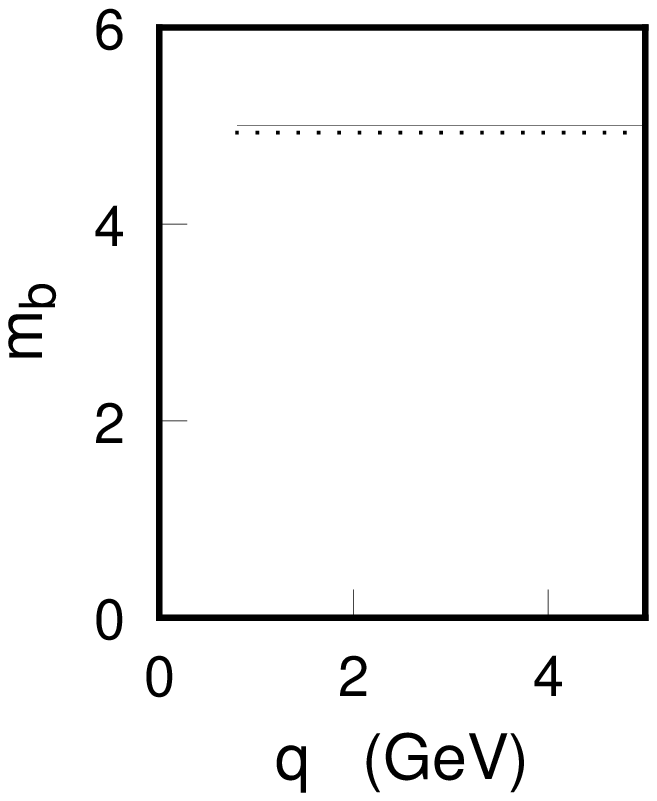}}
\caption{Expected running of the strong coupling constant (left) and the $b$
quark mass (right)
between the energy scale of
$\Upsilon$ physics (around 600-1100 MeV) and the lattice spacing scale
$\pi/a$ at which the coupling is extracted (around 3-6 GeV).
Dotted lines are the quenched approximation, solid lines are full QCD.
}\label{ups}
\end{figure}

If the calculations are done in the quenched approximation, 
one must also estimate the associated corrections and uncertainties.
The expected effects in the $\Upsilon$ meson are illustrated in Fig.~\ref{ups}
Omission of quark loops from the theory results in too large a $\beta$ 
function.  If parameters are set by adjusting middle distance physics such as
the 1P-1S splitting to be right, the coupling constant at short 
distances will be slightly too small, as illustrated in the left hand
figure.
The size of the effect may be
estimated with the aid of potential models
in advance of including sea quarks from  first principles.
  Very roughly, the expected size of the
effect is of order
$\beta_0^{n_f=3}/\beta_0^{n_f=0} = 9/11$.
The ``running mass'' of the $b$ quark, 
on the other hand, does not run for $q^2<m_b^2$.
The effective mass determined at the lattice spacing scale is the same mass
governing the dynamics at the much lower scale of $\Upsilon$ physics,
whether or not the effects of light quark loops are included.
This fact makes $m_b$ the most reliably known of the parameters of QCD.
The light quark masses are more difficult, since we have no way of reliably
estimating the effects of quark loops without including them from first
principles.

\subsubsection*{The Strong Coupling Constant}
To obtain the the lattice scale in physical units for use in 
a lattice determination of the
strong coupling constant,
spin-averaged level splittings in the $\psi$ and $\Upsilon$ systems 
are particularly useful quantities, for reasons already discussed.
Uncertainties arising from usual lattice errors such as finite $V$, 
finite $a$, and statistics are quite small and seem to be under
very good control.
The dominant uncertainties in $\alpha_s$ determinations are arise from
perturbation theory, and, for the present, the quenched approximation.

There are very large perturbative
one loop corrections in the relation between the bare lattice coupling constant
and physically defined coupling constants.
These, however,  are no more significant than the
large corrections in the relation between the $MS$ and
$\overline{MS}$ couplings.  \cite{LM}
The most sensible expansion parameter need not
be the one which is simplest in terms of the regulator:  it must be 
determined from physical quantities.
The bare coupling constants may be rather pathological expansion
parameters.
The origin of the large corrections can also be understood:
they arise from tadpoles due to higher dimension operators 
in the Wilson lattice action.
If such effects are taken into account with mean field theory estimates,
a large number of perturbative series for pure gauge theory, Wilson
fermions, and NRQCD become very convergent.
Mean field theory improvement may also be used be used to estimate higher
order uncalculated terms in lattice perturbation theory.

In Ref. \cite{elk92}, a mean field  improved perturbation theory
was used to extract
the renormalized coupling from the bare lattice coupling.
Subsequent nonperturbative extractions of $\alpha_s$ from a variety of lattice
quantities yielded results which were systematically a few per cent higher 
than the coupling obtained through the mean field improved relation 
with the bare coupling.  \cite{LM,lue93} (See also \cite{mic93}.)
This has resulted in a small but significant increase in the present values
of $\alpha_s$ over those reported in Ref.~\cite{elk92}.

The largest source of uncertainty in current determinations of $\alpha_s$, 
and the one under poorest control is the use of potential models and
perturbation theory to estimate the effects of light quark loops on
the results.  
It is clear that the quenched theory, with too strong a $\beta$ function,
ought to have too weak an $\alpha_s$ at short distances.
For the $\Upsilon$ system, potential models and ordinary perturbation theory
yield similar estimates for the increase in the running for the
quenched theory between the scale of $\Upsilon$ physics and the cutoff scale.
(Potential models suggest 600--1100 MeV for typical gluon momentum transfers
in the $\Upsilon$. \cite{NRQ93})
The agreement of the much more sensitive correction for the $\psi$ system
provides some check on the consistency of the analysis.

The first checks of these estimates from first principles have now appeared.
\cite{ono93}.  With large errors, the results so far are consistent
with the quenched analysis.

The preliminary results
 for the latest analyses of $\alpha_s$ determined in the 
quenched approximation \cite{elk93,NRQ93}
are consistent with
\begin{equation}
  \alpha_s(M_Z) = 0.110 \pm 0.004.
\end{equation}

\subsubsection*{The Heavy Quark Masses}
Since the effective mass of fermions does not run at energy scales below
the pole mass, determinations of $m_c$ and $m_b$ do not suffer from the largest
effect due to the quenched approximation in the determinations of 
$\alpha_s$ and $m_l$: the running of the effective coupling or mass between
the physics scale and the short distance scale at which the coupling is
determined.
The NRQCD collaboration has determined $m_b$ in two independent ways,
with compatible results.  In
method 1, one calculates  the binding energy,
 the difference between the bare quark
masses and the physical mass of the $\Upsilon$.  This lattice result is
then subtracted from the experimental mass of the $\Upsilon$ to
obtain $2m_b$.
In method 2, one
determines the coefficient $1/2m$ required in the quark's 
kinetic energy to obtain the correct energy--momentum
relation for the $\Upsilon$.
The largest uncertainty in each case arises from perturbation theory,
estimated at 1--2\% for method 1, and 2--3\% for method 2.
Uncertainties arising from finite lattice spacing, finite volume,
and statistics are estimated to be 1\% or less.
The result for the pole mass is
\begin{equation}
  m_b = 4.7 \pm 0.1 {\ \rm GeV}.
\end{equation}
Using \cite{mpole}
\begin{equation}
m_{\overline{MS}}(\mu=m)=m_{\rm pole}/(1+4/3 \frac{\alpha_s(m)}{\pi}+12.4 
(\frac{\alpha_s(m)}{\pi})^2+\ldots)  
\end{equation}
and $\alpha_{\overline{MS}}(4.7\ {\rm GeV}) \approx 0.18$ one obtains
\begin{equation}
m_{\overline{MS}}(4.7 \ {\rm GeV})  = 4.2 \pm 0.1 {\ \rm GeV }  
\end{equation}
for the $b$ quark.

The mass of the $c$ quark will soon be reported based on these same 
methods.   It will be relatively less accurate then $m_b$ because the 
errors arising from an additive mass renormalization are larger relative
to $m_c$ then to $m_b$,
and because the required perturbation theory is less accurate at
$\psi$ energy scales (under 700 MeV, according to potential models)
 then at $\Upsilon$ energy scales
(under 600-1100 MeV).

The $t$ quark is expected to decay before it forms QCD bound states.
Lattice methods are unlikely to contribute to determining its mass.

\subsubsection*{The Light Quark Masses}
The light quark masses are the most difficult of the fundamental parameters
of QCD to determine.  There is certainly significant running of the quark
masses between the lattice spacing scale and the scale of light hadron 
physics.  However, there is no hope of estimating the effects of
light quark loops on this running at large distances:  unquenched
calculations are required from the start.
However, $m_\pi$,  the obvious choice to determine 
$m_l$,
is by far the easiest of the light hadron masses to determine,
so this calculation is likely to be among the first to performed reliably
in unquenched calculations.
The current status of determinations of $m_l$ in the $\overline{MS}$ scheme
is summarized in
 Fig.~\ref{figmq} \cite{uka92}.
Calculations have been performed with Wilson fermions and
staggered (KS, or Kogut-Susskind) fermions,
in the quenched approximation ($N_f=0$) and with two flavors of
light quarks ($N_f=2$).
The calculations employing staggered fermions are nicely independent of the 
lattice spacing, while those using Wilson fermions show significant variation,
seeming to approach the staggered results as the lattice spacing is
reduced.  (This may be related to the fact that the lattice spacing
errors in the quark propagators start at $O(a)$ for Wilson fermions
and $O(a^2)$ for staggered.)
This is a pity, because the perturbation theory for the 
relation between the bare staggered fermion mass and the $\overline{MS}$ mass
is much worse behaved than the analogous
perturbation theory Wilson fermions or NRQCD.
There is roughly a 40\% effect in this relation which has not been
understood in terms of  either renormalization group logarithms or
mean field theory tadpoles.

\begin{figure} 
\vspace*{\fl}
\caption{Lattice determinations of the light quark mass in MeV,
  as a function of $\beta\equiv 6/g^2$.
Smaller lattice spacings (i. e., more correct results) are to the right.
}
\label{figmq}
\end{figure}

Current results from staggered fermions
( $m_l\sim 2$ MeV)  are at the low end
of the conventional range, but it is not known yet how reliable 
these are.

There are thus several ways in which the determination of the light quark
masses is more difficult than 
the determination of $\alpha_s$ and the heavy quark
masses:  unquenched calculations are required from the start,
the perturbation theory is less well understood, and 
nonperturbative techniques for extracting the short distance quark mass
are less well developed.
On the other hand, $m_l$ is known from existing phenomenology to within
only a factor of three (as opposed to 5--20\% for the other quantities)
so the payoff will ultimately be bigger.

\section*{Weak Matrix Elements}

As long as it is the case 
that the only clues available to us
 about the theory lying behind the standard model
are the apparently arbitrary ``fundamental'' parameters of the model,
one of the most important applications of lattice gauge theory 
will be the calculations of the hadronic weak matrix elements that
allow the extraction of the elements of the Cabibbo-Kobayashi-Moskawa
matrix elements from hadron decay data.
The hadronic matrix elements for extracting $V_{ud}$ and $V_{us}$ 
can be estimated with sufficient accuracy by employing $SU(2)$ and $SU(3)$
flavor symmetry, respectively, so that lattice calculations are unlikely to
be of much assistance until they are much more accurate.
For the remaining CKM matrix elements, lattice calculations will eventually
play a crucial role.
For the elements connecting $b$ and $c$ quarks for lighter quarks, 
exclusive semileptonic meson decays are the most feasible lattice
calculations.
Cabibbo suppressed semileptonic decays of the $t$ quark are not likely to
be observed any time soon.  The indirect effects of the $t$ quark in
neutral meson mixing amplitudes are the likeliest sources of 
information on CKM matrix elements involving the $t$ quark.
We will discuss some of these amplitudes in order of increasing difficulty.

\subsection*{$B_K$}
The simplest and best understood of these
weak matrix elements is the kaon ``$B$ parameter'',
\begin{equation}\label{B-K}
B_K=\frac{\langle\bar{K}^0|{\cal O}_{\Delta S=2}|K^0\rangle}%
{\frac{8}{3}m_K^2 f_K^2},
\end{equation}
which is required to relate CP violation in kaons to the parameters of the
CKM matrix.
There is  a variety of reasons for this.
\begin{itemize}
 \item The amplitude involves only pseudo scalars, which have the best 
   statistics and finite volume errors of the light hadrons.
 \item Calculations for kaons may be performed with $m_l=m_s/2$ and  
   extrapolated in $m_s-m_d$ to the physical values.  Chiral 
   perturbation theory shows this to be a more benign extrapolation
   then the usual $m_d \rightarrow 0$ limit.  (It is a higher order effect in
   chiral perturbation theory.)
  \item $B_K$ is a ratio of two very similar amplitudes, in which many
   errors (such as those arising from perturbation theory and the quenched
   approximation are likely to cancel.
\end{itemize}
As is typical in calculations with light pseudoscalar mesons,
the cleanest results are produced with staggered fermions, which
preserve an exact chiral symmetry at the expense of ``doubling'' of
the light flavors. \cite{Kil90}

Many of the assumptions in the calculation have been checked in the last 
year.
The one loop perturbative corrections have been checked. \cite{ish93a}
The independence of the result on the use of gauge variant vs. gauge invariant
operators has been tested. \cite{ish93b}
The hope that the effects of the quenched approximation are small has been
tested explicitly. \cite{ish93b,kil93}

\begin{figure} 
\epsfxsize=   \textwidth 
\epsfbox{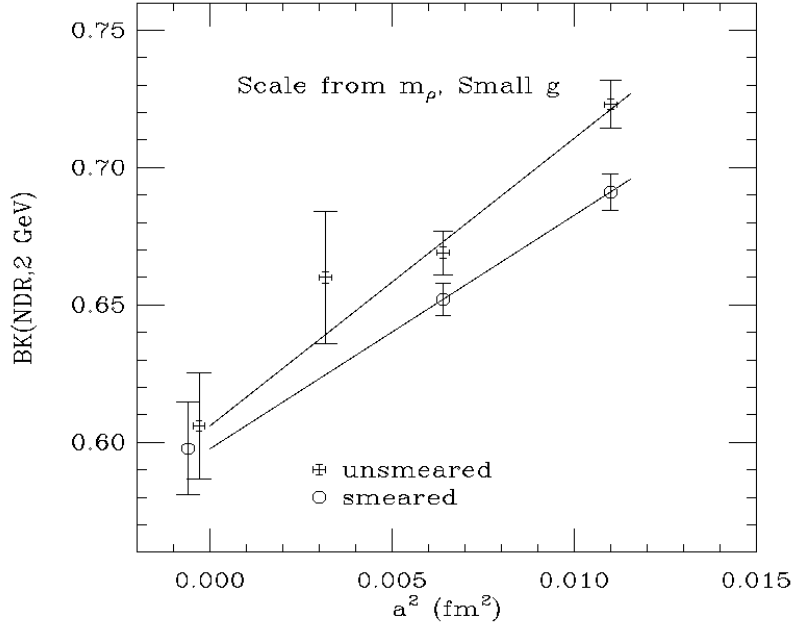}
\caption{$B_K$ in the naive dimensional reduction scheme as a function of the
  lattice spacing squared.  
  Results for two different types of four-quark operators
  have the same $a^2\rightarrow 0$ limit.}\label{figB_K}
\end{figure}

The most important new result is an improved understanding of the 
source of the rather large
finite lattice spacing dependence, which has previously
dominated the uncertainty.
Such lattice spacing dependence in weak amplitudes can arise from
powers of $a$ due to discretization errors,
and from powers of $1/\ln a$ due to perturbation theory.
The perturbative corrections are small, 
mostly canceling between the numerator and 
denominator in Eqn.~\ref{B-K}. They are unlikely to contribute much to 
the finite $a$ effects.  (This is in contrast with some previous examples
of finite $a$ dependence such as in the string tension, where a large
dependence of $\sigma/\Lambda_{lat}^2$ on $a$ is now understood to have arisen
predominantly
from the use of bare lattice perturbation theory rather than renormalized
perturbation theory.)
This leaves the question of the power in $a$ of the effects of 
discretization.
The Staggered Collaboration has very recently completed an examination of
all of the dimension 7 operators capable of producing $O(a)$
errors in $B_K$ for staggered fermions. \cite{sha93}
They have found that among the many such operators, none has the right
flavor and lattice symmetries to contribute to the amplitude for $B_K$.
They therefore extrapolate their small lattice spacing data in
$a^2\rightarrow 0$ (Fig.~\ref{figB_K}) to obtain their final answer.

The current result in the naive dimensional reduction scheme, with estimates
of the known sources of uncertainty, is
\begin{eqnarray*}
B_K(NDR, 2{\rm GeV}) = 0.616 &\pm& 0.020\ (stat) \\
 \mbox{}&\pm& 0.014\ (g^2) \\
 \mbox{}&\pm& 0.009\ (scale) \\
 \mbox{}&\pm& 0.004\ (operator) \\
 \mbox{}&\pm& 0.002\ (correction) \\
=0.616 &\pm& 0.020 \pm 0.017.
\end{eqnarray*}
For the renormalization group invariant $B$ parameter, they obtain
\begin{eqnarray*}
\widehat{B}_K &\equiv& B_K(NDR, 2{\rm GeV})\ \alpha_s(2 {\rm GeV})^{-6/25} \\
&=& 0.825 \pm 0.027 \pm .023.
\end{eqnarray*}

For comparison, the $1/N$ expansion predicts
$\widehat{B}_K = 0.7 \pm 0.1$. \cite{buras93}
A further check which has yet to be done is to test explicitly whether the
extrapolation  to the physical value of
$m_s-m_d$ is as small when quark loops are 
included as it has been shown to be in the quenched approximation.

\subsection*{Heavy Meson Decay Constants}
The special simplicities arising in $B_K$ from the fact that it is a ratio
of two similar amplitudes are not shared by typical weak decay amplitudes
such as heavy meson decay constants.  In particular, there is no reason
to expect perturbative corrections and the effects of the quenched 
approximation to be particularly small.

\begin{figure} 
\epsfxsize=   3in
\hbox{\hspace{.6in} \epsfbox{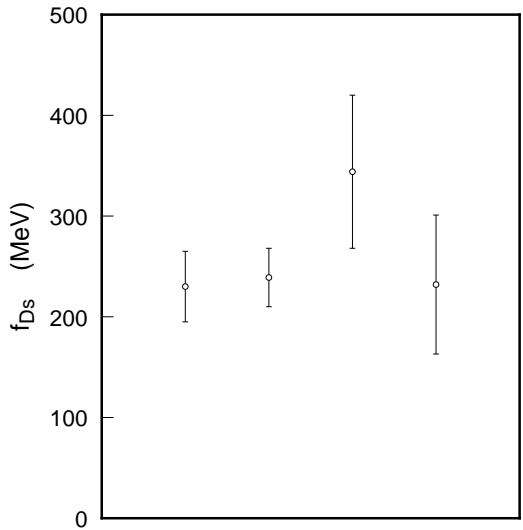}}
\caption{The decay constant of the $D_s$ meson, from lattice calculations
(left two points), compared with experimental results from
   CLEO (third point) and WA75 (last point).
  }  \label{figfDs}
\end{figure}

$f_{D_s}$ is the only heavy meson decay constant which can be directly
compared with experiment.  It will therefore play an important role 
in validating methods for calculating decay constants as theory and experiment
become more precise. Fig.~\ref{figfDs}
shows the lattice results for this quantity, using ``improved'' (in terms
of finite lattice spacing errors) light quark methods. They are
$f_{D_s} = 230\pm 35$ MeV \cite{fDs1} and $218^{+50}_{-8}$ MeV \cite{fDs2}.
They may be compared 
with the experimental numbers $232\pm 69$ MeV from WA75\cite{fDs3} and 
$344\pm 76$ MeV from CLEO \cite{fDs4}.
The analysis going into the lattice uncertainties is not as detailed as 
that behind $B_K$.
The numbers have remained reasonably stable, however,  as the calculational
methods have improved over the last few years.

The $B$ meson decay constant $f_B$ is of even greater phenomenological interest
because of its role in describing $B^0 \overline{B^0}$ mixing, 
and much work has been invested in it recently.
Initial lattice calculations in the static limit
produced results which were very high
(over 300 MeV), compared with expectations from quark model 
estimates.~\cite{Eic90}
Subsequent work  revealed several sources of mostly negative  corrections,
and a final consensus has not yet emerged, even in the quenched 
approximation and in the static limit.
Current estimates range from 185 to 370 MeV.~\cite{fDs1,fDs2,Ale92,All93,Eic93}

\begin{figure} 
\epsfxsize=   \textwidth 
\epsfbox{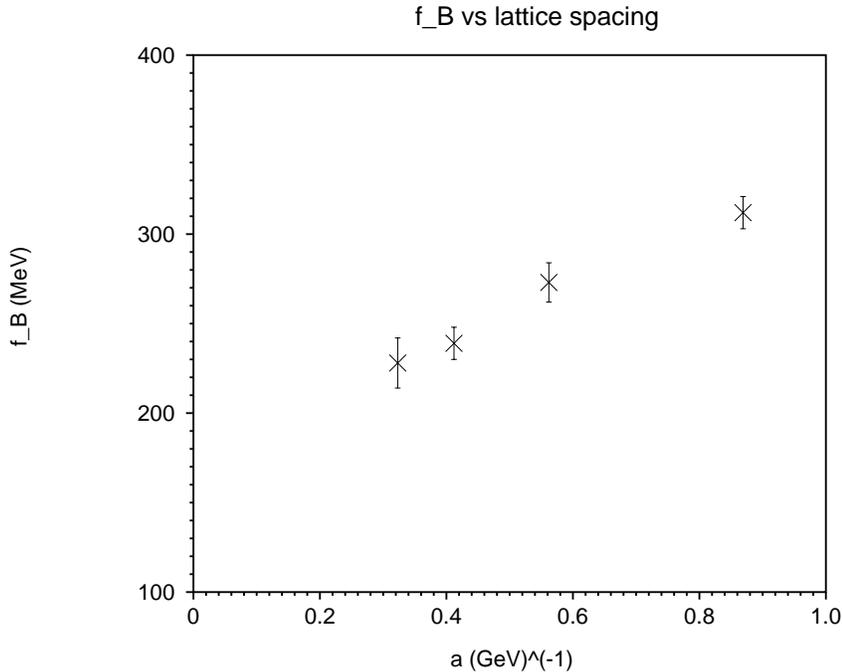}
\caption{The decay constant of the $B$ meson, $f_B$, in the static 
  approximation, as a function of the lattice spacing $a$.
  }  \label{figfB}
\end{figure}

An example of a correction which is still in the process of being sorted out
is shown in Fig.~\ref{figfB}.~\cite{Eic93}
More dependence on the lattice spacing is observed than was apparent
originally.  \cite{Ale92,Eic93}
Part of this $a$ dependence arises from higher orders of perturbation theory
which fall like logarithms of $a$ as $a\rightarrow 0$,
and can be at least partially ameliorated with the use of improved
perturbation theory.
On the other hand, part of it may also come from discretization errors
which fall as powers of $a$.
Until the functional form of the $a$ dependence is understood, 
an $a\rightarrow 0$ extrapolation cannot be made with confidence.

Ratios of decay constants can be calculated more accurately.
Current results for $f_{B_s} /f_B $ ($=f_{D_s} /f_D(1+O(m_s/m_c) $
\cite{Gri93}) in the quenched approximation lie in the range 
1.11--1.22.~\cite{fDs1,fDs2,Ale92,All93}
The effects of adding quark loops to these calculations may be estimated
from the one loop chiral perturbation theory calculations of these quantities,
which give $f_{B_s} /f_B =f_{D_s} /f_D \approx 1.1$.~\cite{Gri92}
Ultimately, on would hope to calculate the deviations of these ratios
from unity as  accurately as the decay constants themselves:
that is, $f_{B_s} /f_B-1,$ $ f_{D_s} /f_D-1$, and eventually 
$f_{B} /f_D-1$ to perhaps 20\%.

The hadronic amplitude for $B^0 \overline{B^0}$ mixing may
be written in terms of $f_B$ and a $B$ parameter, following the notation
of the $K$ system.
In the standard model, the parameter measuring the
experimentally observed $B^0 \overline{B^0}$ mixing is given by
\begin{equation}
x_d = (\mbox{known factors})\,|V_{td}^*V_{tb}|^2 f_B^2 B_B.
\end{equation}
Pilot studies of $B_B$ and $B_{B_s}$ have been performed which yielded
results close to the ``vacuum saturation'' value of one.
 \cite{Aba92}

\subsection*{Semileptonic Decays}
Semileptonic decay amplitudes share all of the difficulties of decay constants.
In addition,
\begin{itemize}
  \item   large momentum in the decay meson leads to worse 
    finite lattice spacing  errors   and worse statistics,
  \item finite lattice volume leads to a coarse decay momentum discretization, 
    and
  \item calculations for many decay momenta are required, each of which is
    as difficult as a decay constant calculation.
\end{itemize}

Although lattice calculations are first principles calculations, 
it is perhaps fair they are 
treated in competition with sum rules and
quark models at the present stage of the game.  \cite{Wit93}
Getting the  level of detail and the accuracy of the
uncertainty analysis for these processes 
to match that already obtained
for $B_K$ will be a long  process, even if all
goes relatively well.
However, there is no obstacle presently known to eventually getting 
semileptonic decays into comparably good shape (other than the 
requirement of doing a lot of  work).

Most lattice work on this area in the last two years has focused on
the calculation of the Isgur-Wise function $\xi(v\cdot v')$
on the lattice.
Two approaches have been investigated:
\begin{enumerate}
\item  Direct formulation of an action for quarks in
  the $m\rightarrow \infty$ limit for finite
  velocity.  \cite{man92}  (This is analogous to the static approximation for
  $v=0$.)
\item Use of light quark methods to calculate $D$ meson elastic 
  scattering \cite{ber93,boo93}, using
   \begin{equation}
     \langle D_{v'}| \overline{c} \gamma_\mu c| d_v\rangle = 
       \xi(v\cdot v')  (p +p')_\mu+O(\Lambda/m_c).
   \end{equation}
\end{enumerate}

\begin{figure} 
\epsfxsize=  3in
\vskip-.9in
\hbox{\hskip.9in \epsfbox{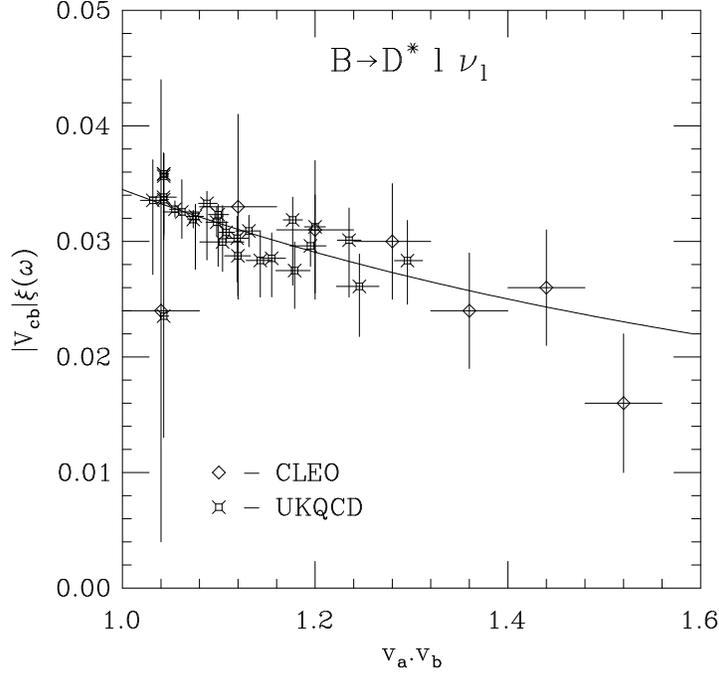} }
\vskip.6in
\caption{The Isgur-Wise function calculated on the lattice 
    compared with $B$ decay data from CLEO.
  }  \label{figIW}
\end{figure}
 
Fig.~\ref{figIW} shows the Isgur-Wise function calculated in the second
approach in Ref.~\cite{boo93}, compared with $B$ decay data from CLEO.
The lattice error bars do not explicitly include uncertainties arising
from finite lattice spacing, finite volume, the quenched 
approximation, or $1/m_c$ effects.  $\xi(0)=0$ has been used 
to normalize the lattice results. The lattice calculations are not
yet accurate enough to determine the curvature of the function, so
 what is really being calculated is the slope.
Using the Stech-Neubert-Rieckert parameterization of the function,
the lattice groups quote for the shape parameter
$\rho^2= 1.41 (0.19) (0.19)  $ \cite{ber93}, and
$\rho^2= 1.6^{+7}_{-6}  $ \cite{boo93},
which are compatible results from sum rules and from fitting the shape 
of the data directly.
The values of $|V_{cb}|$ obtained by the lattice groups 
are therefore compatible with those obtained from other analyses.
Normalizing to a $B$ lifetime of 1.50 ps, they obtain
$|V_{cb}|\sqrt{\tau_B/1.50 {\rm ps}} = 0.044$ in Ref.~\cite{ber93} and
$0.043 (2) (^6_5)$ in Ref.~\cite{boo93}.
In the errors quoted in Ref.~\cite{boo93}, the first error is
experimental, the second is part of the theoretical uncertainty.

\section*{Conclusions}
There are now several lattice calculations ($B_K$, $m_b$, $\alpha_s$)
for which at least a first attempt has been made to examine
all of the largest sources of uncertainty quantitatively.
These uncertainty estimates are not yet on a par with the analysis
of $g-2$ for the electron (although eventually they should be),
but they are quite competitive with the analysis of theoretical uncertainties
in short distance perturbative QCD processes.

The calculations which are currently best understood
are in one way or another special cases, simpler than the generic
lattice calculation.
However, they and many others of the most interesting 
phenomenological calculations (decay constants, $B$ parameters,
many semileptonic decays)
share certain other simplicities,
which put them into a class which is likely to be
doable
over the next few years, assuming only programmatic rather than
revolutionary improvements in methods.
They involve hadronically stable mesons, either pseudoscalars or
heavy quark-antiquark.  They involve processes with a single
hadron existing at a time.

Baryons and unstable mesons are likely to prove a bit more
demanding, though still well within the range of current methods.
More demanding still will be processes involving more than one hadron.
(Conceptual problems involving final state interactions in imaginary
time, as is used in lattice calculations,
 have yet to be worked out in practical applications.)
The most phenomenologically  important of these are the hadronic
kaon decay amplitudes necessary for the analysis of CP violation
in the $K$ system.
Farther away still are such things as a full nonperturbative calculation
of high energy $PP$ scattering, which are certainly not immediate prospects.
Setting our sights even higher, one would like eventually
to have lattice methods that worked for chiral gauge theories,
  so that nonperturbative
beyond the standard model physics could be investigated in a reliable
and straightforward way.
No proposed method for such theories has so far been proven to work.
 It is not yet known whether this is a result of simple 
technical difficulties which are unusually complicated, or
whether it is an indication of something deep about these theories
which has not yet been sufficiently appreciated.

There are many goals  ahead of us, not all of which are yet
within either our grasp or our  reach.
On the other hand, most (though not all) of the calculations
which are most crucial in extracting the fundamental parameters of the
standard model from experiment are in the simplest class of lattice
calculations:  they involve single, stable mesons.
The simplest of these have now been completed with uncertainty
estimates.  There is a good hope that these uncertainty estimates
 can be made very solid, and that many more simple but 
important calculations will join them over the next few years. 

\section*{Acknowledgments}
I would like to thank many of the authors of work discussed here
for correspondence.
I would also like to thank C. Bernard, S. Gottlieb, 
A. S. Kronfeld, G. P. Lepage,  G. Martinelli,
   S. Sharpe,   D. Toussaint, 
and A. Ukawa 
for helpful discussions.

Fermilab is operated by Universities Research Association under contract
with the U. S. Department of Energy.


\begin{references}
\bibitem{confs} For a more general and detailed  overview of the 
 current status of lattice gauge theory, one may consult recent proceedings
 of the annual lattice conference, for example Refs. \cite{LAT90},
 \cite{LAT91}, and  \cite{LAT92}.
 Some of the material in this review is discussed in more detail in 
 Ref.~\cite{annrev}.
\end{references}
\end{document}